\documentclass[letterpaper, 10 pt, conference]{ieeeconf}  % Comment this line out
     
\usepackage{graphicx}
\usepackage[separate-uncertainty=true]{siunitx}
\usepackage{amsmath}

\IEEEoverridecommandlockouts                              % This command is only
\usepackage{xr}

\title{\LARGE \bf
Oscillatory fluid flow drives scaling of contraction wave with system size
}

\author{Jean-Daniel Julien$^{1,*}$ and Karen Alim$^{1}$% <-this % stops a space
\thanks{$^{1}$Max Planck Institute for Dynamics and Self-Organization,
        G\"ottingen, Germany
        }%
}

\begin{document}

\maketitle
\thispagestyle{empty}
\pagestyle{empty}

%%%%%%%%%%%%%%%%%%%%%%%%%%%%%%%%%%%%%%%%%%%%%%%%%%%%%%%%%%%%%%%%%%%%%%%%%%%%%%%%
\begin{abstract}

Flows over remarkably long distances are crucial to the functioning of many organisms, across all kingdoms of life. Coordinated flows are fundamental to power deformations, required for migration or development, or to spread resources and signals. A ubiquitous mechanism to generate flows, particularly prominent in animals and amoeba, is acto-myosin cortex driven mechanical deformations that pump the fluid enclosed by the cortex. Yet, it is unclear how cortex dynamics can self-organize to give rise to coordinated flows across the largely varying scales of biological systems. Here, we develop a mechanochemical model of acto-myosin cortex mechanics coupled to a contraction-triggering, soluble chemical. The chemical itself is advected with the flows generated by the cortex driven deformations of the tubular-shaped cell. The theoretical model predicts a dynamic instability giving rise to stable patterns of cortex contraction waves and oscillatory flows. Surprisingly, simulated patterns extend beyond the intrinsic length scale of the dynamic instability - scaling with system size instead. Patterns appear randomly but can be robustly generated in a growing system or by flow-generating boundary conditions. We identify oscillatory flows as the key for the scaling of contraction waves with system size. Our work shows the importance of active flows in biophysical models of patterning, not only as a regulating input or an emergent output, but rather as a full part of a self-organized machinery. Contractions and fluid flows are observed in all kinds of organisms, so this concept is likely to be relevant for a broad class of systems.

\end{abstract}

%%%%%%%%%%%%%%%%%%%%%%%%%%%%%%%%%%%%%%%%%%%%%%%%%%%%%%%%%%%%%%%%%%%%%%%%%%%%%%%%
\section*{Introduction}
%%%%%%%%%%%%%%%%%%%%%%%%%%%%%%%%%%%%%%%%%%%%%%%%%%%%%%%%%%%%%%%%%%%%%%%%%%%%%%%%

Fluid flows are fundamental to the functioning of all organisms. They play an important role in homeostasis, by spreading resources and biochemical signals %over scales too large for diffusion to be reliable 
\cite{goldstein_physical_2015,alim_mechanism_2017,koslover_cytoplasmic_2017}. 
They power deformations driving migration of many motile cells \cite{allen_cytoplasmic_1978,blaser_migration_2006,rieu_periodic_2015}, and can even directly impact on organism size \cite{tominaga_cytoplasmic_2013}.
Surprisingly, even in the absence of a pace-maker like a heart, flows are coordinated on vastly different scales ranging from the size of a single migrating cell of about \SI{20}{\mu m} \cite{Yoshida:2006,Lammermann:2008}, via the \textit{C.~elegans} gonad of about \SI{450}{\micro\meter} \cite{wolke_actin-dependent_2007,Atwell:2015}, the \textit{Drosophila} embryos of about \SI{500}{\micro\meter} \cite{hecht_determining_2009}, to acellular slime molds of more than \SI{2}{\centi\meter} in size \cite{alim_random_2013}. The physical mechanism of how coordinated flows can self-organize particularly in a single cellular envelope remains unknown. 

Animal and slime mold cells are lined with an acto-myosin cortex situated just below their cellular envelope, enclosing the cells fluid cytoplasm. This acto-myosin mesh-work forms an active visco-elastic material \cite{kruse_generic_2005,prost_active_2015}.
It contracts under myosin motor activity, and thereby drives the enclosed cytoplasm to flow into a less contracted part of the cell \cite{Taylor:1973,goldstein_physical_2015}. Long-ranged flows therefore require a spatial organization of cortex contractility. 
While the mechanism underlying long-ranged flows driven by the entrainment of fluid with transported vesicles has been studied in depth \cite{Goldstein:08,tominaga_molecular_2015,quinlan_cytoplasmic_2016},
the mechanism driving the coordination of cortex contraction across a cell or an organism is unclear.

In migrating cells or amoeba the cortex exhibits contractions driving oscillatory fluid flows and migration of the tubular shaped cell \cite{Lammermann:2008,lewis_coordination_2015}.
Already in the early 1980's Oster and Odell explored the idea that a contraction-triggering chemical, like calcium, could explain dynamic, oscillatory patterns of acto-myosin activity in the cell cortex \cite{oster_mechanics_1984,oster_mechanochemical_1984}. Yet, the dynamics' spatial component was not investigated.
Calcium is necessary to generate acto-myosin contractions in many biological systems \cite{smith_model_1992,schuster_modelling_2002,levasseur_novel_2007,antunes_coordinated_2013}. Additionally, calcium is regulated by mechanical stretching, via mechanosensitive channels  \cite{glogauer_calcium_1997,lee_regulation_1999,matthews_cellular_2006,vogel_local_2006}.
Consequently, cortex expansion triggers the influx of calcium which in turn leads to contraction.
Due to the widespread importance of cortex activity in developmental processes, this feedback is of general interest, investigated in mechanochemical models \cite{veksler_calcium-actin_2009}.
Models describing the cortex as a fluid \cite{Bois:2011} or as a poroelastic medium \cite{Radszuweit:2013}, where active stress is up-regulated by a chemical immersed in this medium, account for short-ranged traveling waves of contractions and oscillatory flows. Yet, the mechanistic insight is missing that can account for coordination of contractions on scales beyond the intrinsic length scale of the dynamic system and thus account for very long-ranged fluid flows.

%%%%%%%%%%%%%%%%%%%%%%%%%%%%%%%%%%%%%%%%%%%%%%%%%%%%%%%%%%%%%%%%%%%%%%%%%%%%%%%%
\begin{figure*}[t]
\centering
\includegraphics[width=.95\linewidth]{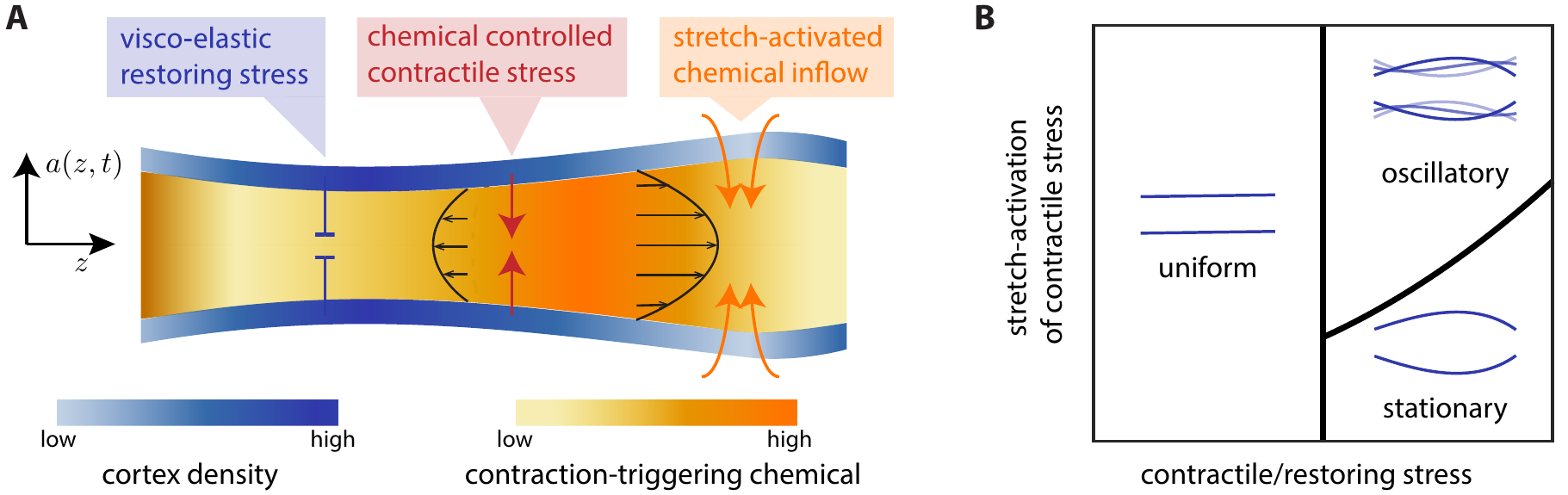}
\caption{Illustration of the model predicting self-sustained contraction waves. (A) A tubular shaped cell lined with an acto-myosin cortex (blue) enclosing the liquid cytoplasm carrying a contraction-triggering chemical (orange). Chemical controlled contractile cortex stress is balanced by visco-elastic restoring stress. Small cortex contractions self-amplify as more actin overlaps in a contracted cortex, see variations in cortex density. Cortex stretch leads to inflow of contraction-triggering chemical allowing for self-sustained oscillations. Contractions are coupled spatially as the chemical is advected with the cytoplasmic flows (parabolic lines) resulting from cortex deformations. (B) Phase diagram depicting region of uniform, stationary pattern and oscillatory patterns.}
\label{fig:scheme}
\end{figure*}
%%%%%%%%%%%%%%%%%%%%%%%%%%%%%%%%%%%%%%%%%%%%%%%%%%%%%%%%%%%%%%%%%%%%%%%%%%%%%%%%
Particularly, the slime mold \textit{Physarum polycephalum} is renown for long-ranged coordination of cortex contractions. Here, fluid flows scale with organism size from \SI{2}{mm} to at least \SI{2}{cm} \cite{alim_random_2013}. Again, flows are known to power organism migration \cite{rieu_periodic_2015,lewis_coordination_2015}. Moreover, stimulants that alter cortex contractility have recently been found to be advected with the fluid flows inside the cell \cite{alim_mechanism_2017}. This observation suggests that the physical transport by fluid flows is the key to long-ranged spatial coordination of cortex contractions and fluid flows.

Here, we investigate in a tubular geometry the self-organization of cortex contractions, coupled to a contraction-triggering chemical which is advected with the flows of the fluid cytoplasm. The simple two-component model is unstable toward self-sustained cortex oscillations as cortex stretching triggers the increase of the contraction-triggering chemical concentration. A linear analysis of the model predicts traveling wave solution of cortex shape and the intrinsic wavelength of the dynamic instability is derived. In contrast to our analytic prediction, numerical solutions of the model in a tube with periodic boundaries show a probabilistic distribution of five different patterns of traveling waves. Although the tube is twice as long as the expected intrinsic wavelength, in one of these patterns the traveling wave scales with tube length. 
Further analysis shows that scaling can be robustly generated in growing tubes with periodic boundary conditions or by flow-generating boundary conditions in non-growing tubes. We identify oscillatory flows as the key to the scaling of contraction waves with system size. The ubiquity of fluid flows in biological and non-living systems suggests that this non-trivial scaling could be broadly relevant in active matter.
%%%%%%%%%%%%%%%%%%%%%%%%%%%%%%%%%%%%%%%%%%%%%%%%%%%%%%%%%%%%%%%%%%%%%%%%%%%%%%%%
\section*{Results}
%%%%%%%%%%%%%%%%%%%%%%%%%%%%%%%%%%%%%%%%%%%%%%%%%%%%%%%%%%%%%%%%%%%%%%%%%%%%%%%%
\subsection*{Coupling of tubular-shaped cortex with contraction triggering chemical}
%%%%%%%%%%%%%%%%%%%%%%%%%%%%%%%%%%%%%%%%%%%%%%%%%%%%%%%%%%%%%%%%%%%%%%%%%%%%%%%%
A cell showing coordinated cytoplasmic fluid flows in general has a distinctive viscous fluid phase separated from a surrounding visco-elastic acto-myosin cortex. The nature of long-ranged flows typically entails a tubular cell shape. We here consider as a minimal model an active, visco-elastic tube of length $L$,  filled with a fluid. Tube shape is fully defined by the tube's radius $a\left(z,t\right)$ along the tube's axial position $z$ and over time $t$.
The tube's temporal evolution directly follows from the conservation of the fluid volume within the tube
\begin{equation}
\frac{\partial a^2}{\partial t}=-\frac{\partial}{\partial z}\left(a^{2}\bar{u}\right),
\label{eq:radius}
\end{equation}
where $\bar{u}(z,t)$ denotes the cross-sectionally averaged fluid flow velocity along the tube. 
Fluid flow is powered by contractions of the tube, and thus by the stress $\sigma(z,t)=\sigma_{\mathrm{c}}+\sigma_{\mathrm{e}}$ acting radially within the tube's cross-section, see Fig.~\ref{fig:scheme}. We distinguish, $\sigma_{\mathrm{c}}$, the contractile stress stemming from acto-myosin activity within the cortex and, $\sigma_{\mathrm{e}}$, the counteracting visco-elastic restoring stress of the cell. For long slender tubes, $a/L\ll 1$, Lubrication approximation applies. The Stokes equation for the fluid flow simplifies to
\begin{equation}
\bar{u}=-\frac{a^{2}}{8\mu}\frac{\partial}{\partial z}\left(\sigma_{c}+\sigma_{\mathrm{e}}\right),
\end{equation}
where $\mu$ denotes the dynamic viscosity of the fluid. We approximate the cell's material properties to be dominated by a linear viscoelastic response with a small non-linearity to suppress potential divergences. 
Abbreviating radial deformation as $\epsilon=[a-a^{*}]/a^{*}$ with respect to the constant equilibrium radius $a^{*}$, the restoring stress is given by
\begin{equation}
\sigma_{\mathrm{e}}=E\epsilon+\kappa\epsilon^{3}+\eta\frac{\partial \epsilon}{\partial t},
\end{equation}
where $E$ and $\eta$ denote the tube's effective elastic modulus and viscosity, respectively, and $\kappa$ the strength of the non-linear response. Note that $E$ and $\eta$ incorporate both the elastic properties and the thickness of the cell cortex. 

In light of the role of calcium in coordinating acto-myosin activity, we describe the strength of the cortex contractile stress to be proportional to the concentration of a contraction-triggering chemical $c$. In addition, contractions may self-amplify as more actin-fibers overlap in a contracted cortex. Inversely, overlap decreases in an expanded cortex, reducing potential contractility. Consequently, the contractile stress is represented by: 
\begin{equation}
\sigma_c=\sigma_0\frac{\mathcal{C}}{\mathcal{C}^*}\left(1-\frac{\epsilon}{\epsilon_{\sigma}}\right).
\end{equation}
Here, $\mathcal{C}=\pi a^2 c$ represents the chemical concentration integrated across the cross-section of the tubular cell, $\mathcal{C}^*$ is the equilibrium concentration, $\sigma_0$ describes the strength of active tension at equilibrium, and $\epsilon_{\sigma}$ the typical deformation for the change in fiber overlap to become significant for contractility. 
The chemical itself constantly cycles between an inactive state and an active state in the cytoplasm with release rate $p_c$ and capture rate $d_c$. We assume the amount of inactive chemical to be abundant and therefore not of concern for the dynamics investigated here.
Importantly, motivated by our knowledge on calcium regulation by mechanical deformations \cite{glogauer_calcium_1997, lee_regulation_1999, matthews_cellular_2006,vogel_local_2006},
additional chemical is released at the cell's membrane upon cortex stretch with $\epsilon_c$ denoting the corresponding typical deformation scale.
Now, we further incorporate spatial coupling as we account for the advection and diffusion of the contraction-triggering chemical. Reflecting the tubular cell shape we assume the chemical to average out quickly across the tubes cross-section by diffusion, with diffusivity $D$, compared to the advective transport with velocity $\bar{u}$ along the tube of length $L$, $a^2 \bar{u}/DL\ll 1$. This assumption warrants the use Taylor dispersion for a tube of varying radius \cite{Mercer:90,Mercer:94},
\begin{eqnarray}
\frac{\partial\mathcal{C}}{\partial t} & = & 2\pi a\left[p_{c}\left(1+\frac{\epsilon}{\epsilon_{c}}\right)-d_{c}\frac{\mathcal{C}}{\pi a^{2}}\right]\nonumber\\
 &  & +\frac{\partial}{\partial z}\left[-\mathcal{C}\bar{u}+\left(D+\frac{a^{2}\bar{u}^{2}}{48D}\right)\pi a^{2}\frac{\partial}{\partial z}\left(\frac{\mathcal{C}}{\pi a^{2}}\right)\right].
\label{eq:chemical}
\end{eqnarray}
Note that the $2\pi a$ factor in front of the reaction term comes from the assumption that although the chemical is present in the fluid, its contraction-triggering action takes place at the surface of the tube.
%%%%%%%%%%%%%%%%%%%%%%%%%%%%%%%%%%%%%%%%%%%%%%%%%%%%%%%%%%%%%%%%%%%%%%%%%%%%%%%%
\subsection*{Stretch-activated chemical inflow controls self-sustained oscillations}
%%%%%%%%%%%%%%%%%%%%%%%%%%%%%%%%%%%%%%%%%%%%%%%%%%%%%%%%%%%%%%%%%%%%%%%%%%%%%%%%
At zero fluid flow the tube's radius is uniformly at its rest value of $a=a^{*}$. Similarly, the chemical is at a constant value of $C^{*}=\pi a^{*2} p_c/d_c$ throughout the tube.
From the model setup, we expect this uniform state to be unstable with respect to small perturbations as a small deformation of the tube radius grow when the contractile stress of scale $\sigma_0/\epsilon_{\sigma}$ exceeds restoring stress, see Fig.~\ref{fig:scheme} B. Proportional to the relative stretch parameterized by $\epsilon_{c}$, a stretched cortex additionally ignites the inflow of the contraction-triggering chemical. Chemical inflow results in contractions, thus decreasing the deformation, and initiating the oscillation. Consistent with this intuitive reasoning, linear stability analysis shows that the uniform state is unstable if $\sigma_0/\epsilon_{\sigma}$ is large enough compared to the tube's elastic modulus $E$, diffusion $D$, and capture rate $d_c$, all factors limiting the development of fluctuations,
%\begin{equation}
%\protect\begin{array}{c}
%\alpha=\frac{\sqrt{\frac{\sigma_0}{\epsilon_{\sigma}}-E}}{\sqrt{\frac{16\mu D}{a^{*2}}}+\sqrt{\frac{2d_{c}\eta}{ a^{*}}}}>1,
%\protect\end{array}
%\label{eq:instability condition}
%\end{equation}
\begin{equation}
\protect\begin{array}{c}
\frac{\sqrt{\frac{\sigma_0}{\epsilon_{\sigma}}-E}}{\sqrt{\frac{16\mu D}{a^{*2}}}+\sqrt{\frac{2d_{c}\eta}{ a^{*}}}}>1,
\protect\end{array}
\label{eq:instability condition}
\end{equation}
see SI Text. The wavelength of the most unstable mode is given by 
\begin{equation}
\begin{array}{c}
\lambda_{\mathrm{lin}}=\pi a^{*}\sqrt{ \frac{\eta}{\mu}} \left( \sqrt{ \frac{a^{*2}\left( \frac{\sigma_0}{\epsilon_{\sigma}}-E\right)}{D\mu}}-4\right)^{-\frac{1}{2}}.
\end{array}
\label{eq:most unstable wave number}
\end{equation}
The scale of coordinated flows set by this intrinsic wavelength here arise from the competition between diffusion $D$ and viscosity $\eta$, increasing the wavelength by filtering out perturbations on a short scale, and contractility, amplifying the fluctuations locally and controlled essentially by $\sigma_0/\epsilon_\sigma$.
Approximating $\mu D\ll\frac{\sigma_0a^{*2}}{\epsilon_{\sigma}}-E a^{*2}$, the expression for the intrinsic wavelength for a system with an oscillatory pattern, $\frac{\sigma_0}{\epsilon_\sigma}-E\sim E$, simplifies to $\lambda_{\mathrm{lin}}\simeq\pi\sqrt{a^* \eta}\left[\frac{D}{\mu E}\right]^{1/4}$. Further, an analytical expression for the oscillation frequency $\omega$, see Eq. S1, and the onset of oscillations is tractable.
%\begin{equation}
%\beta=\sqrt{\frac{2\sigma_0}{\epsilon_{c}}\frac{ a^{*}}{\eta d_{c}}}>1+\alpha^{2},
%\label{eq: oscillation condition}
%\end{equation}
\begin{equation}
\sqrt{\frac{2\sigma_0}{\epsilon_{c}}\frac{ a^{*}}{\eta d_{c}}}>1+\left(\frac{\sqrt{\frac{\sigma_0}{\epsilon_{\sigma}}-E}}{\sqrt{\frac{16\mu D}{a^{*2}}}+\sqrt{\frac{2d_{c}\eta}{ a^{*}}}}\right)^{2},
\label{eq: oscillation condition}
\end{equation}
see Fig.~\ref{fig:scheme} B. The result confirms the intuitive idea that oscillations occur if the stretch-activated chemical release, controlled by $1/\epsilon_c$, is strong enough to counterbalance the self-amplifying deformation of the tube. Based on these analytical results we expect the system to generate spontaneous contractile waves of a wave size given by the most unstable mode $\lambda_{\mathrm{lin}}$. 
%%%%%%%%%%%%%%%%%%%%%%%%%%%%%%%%%%%%%%%%%%%%%%%%%%%%%%%%%%%%%%%%%%%%%%%%%%%%%%%%
\subsection*{Multiple patterns of contractions arise in a periodic tube}
%%%%%%%%%%%%%%%%%%%%%%%%%%%%%%%%%%%%%%%%%%%%%%%%%%%%%%%%%%%%%%%%%%%%%%%%%%%%%%%%
\begin{figure}[tbp]
\centering
\includegraphics[width=.95\linewidth]{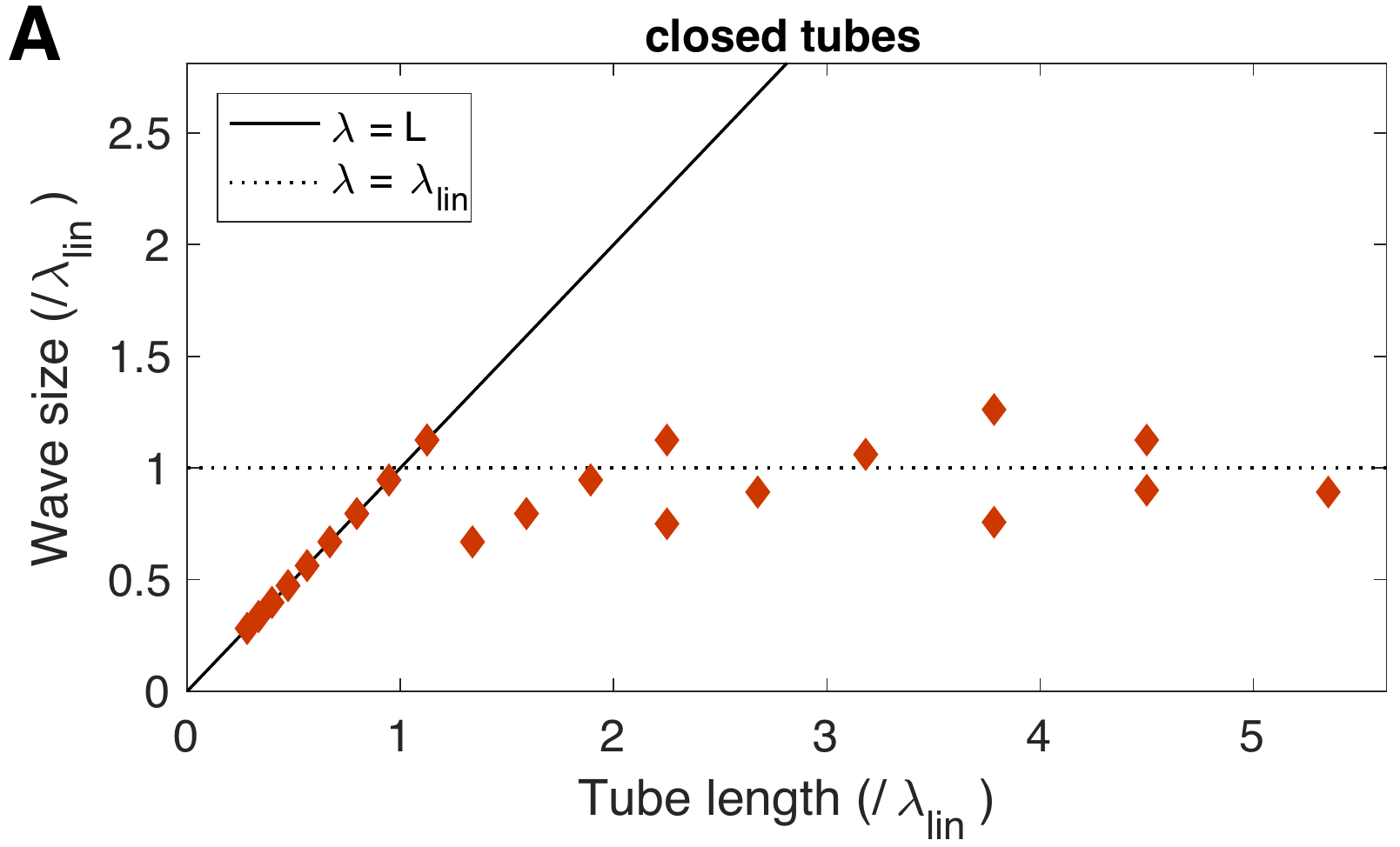}
\centering
\includegraphics[width=.95\linewidth]{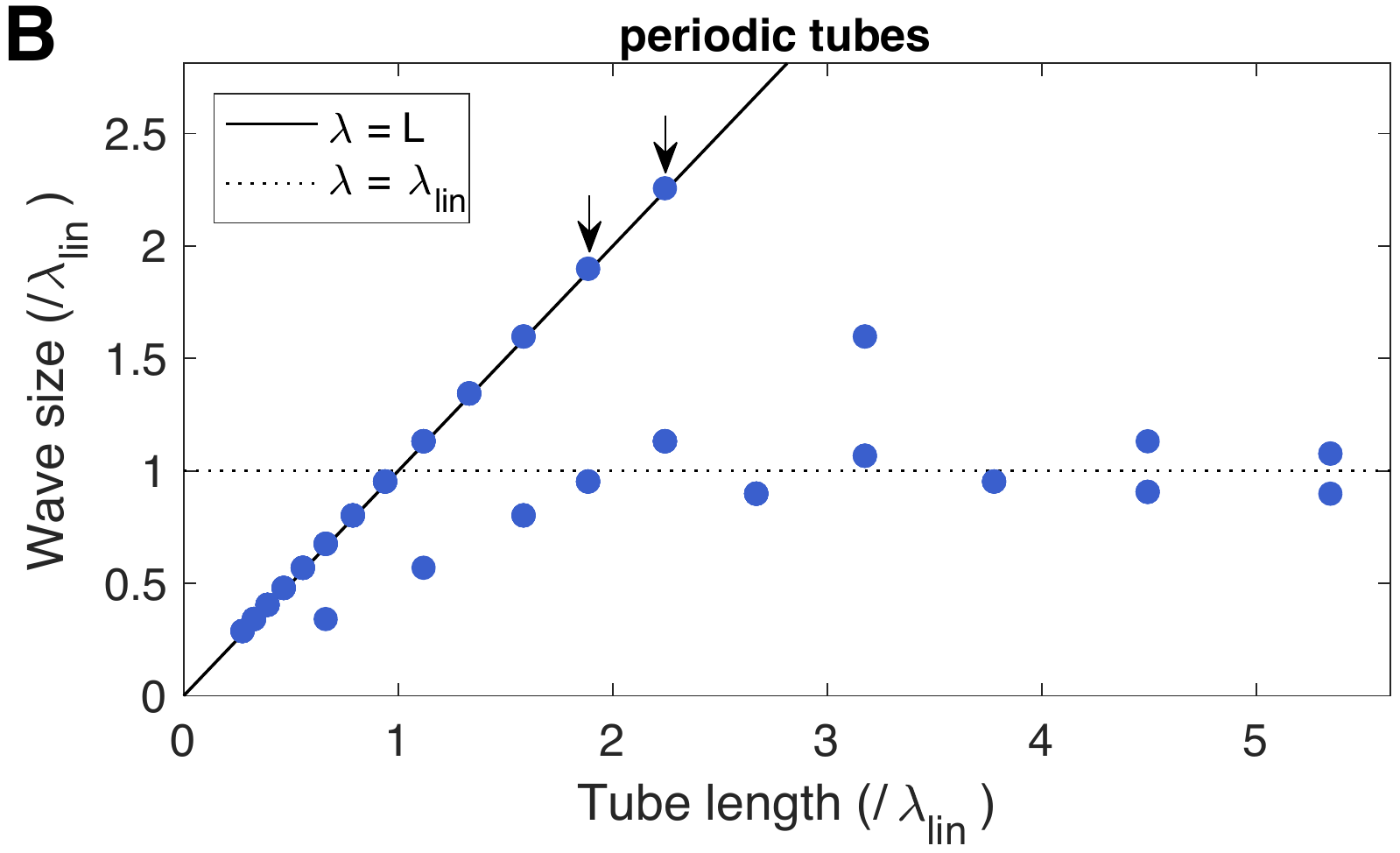}
\caption{Contraction wave may scale with tube length. Size of self-sustained contractions waves $\lambda$ for tubes of different lengths $L$, for closed (A) and periodic (B) boundary conditions. Linear stability analysis predicts wave size to be bound by the intrinsic wavelength of the most unstable mode $\lambda_{\mathrm{lin}}$ (dotted line) as observed numerically for a closed tube. Yet, wave size may scale with tube length (arrows) for periodic boundary conditions. Each data point denotes an individual simulation.}
\label{fig:SimulationNoGrowth}
\end{figure}
%%%%%%%%%%%%%%%%%%%%%%%%%%%%%%%%%%%%%%%%%%%%%%%%%%%%%%%%%%%%%%%%%%%%%%%%%%%%%%%%
%%%%%%%%%%%%%%%%%%%%%%%%%%%%%%%%%%%%%%%%%%%%%%%%%%%%%%%%%%%%%%%%%%%%%%%%%%%%%%%%
\begin{figure*}[t]
\centering
\includegraphics[width=0.85\paperwidth]{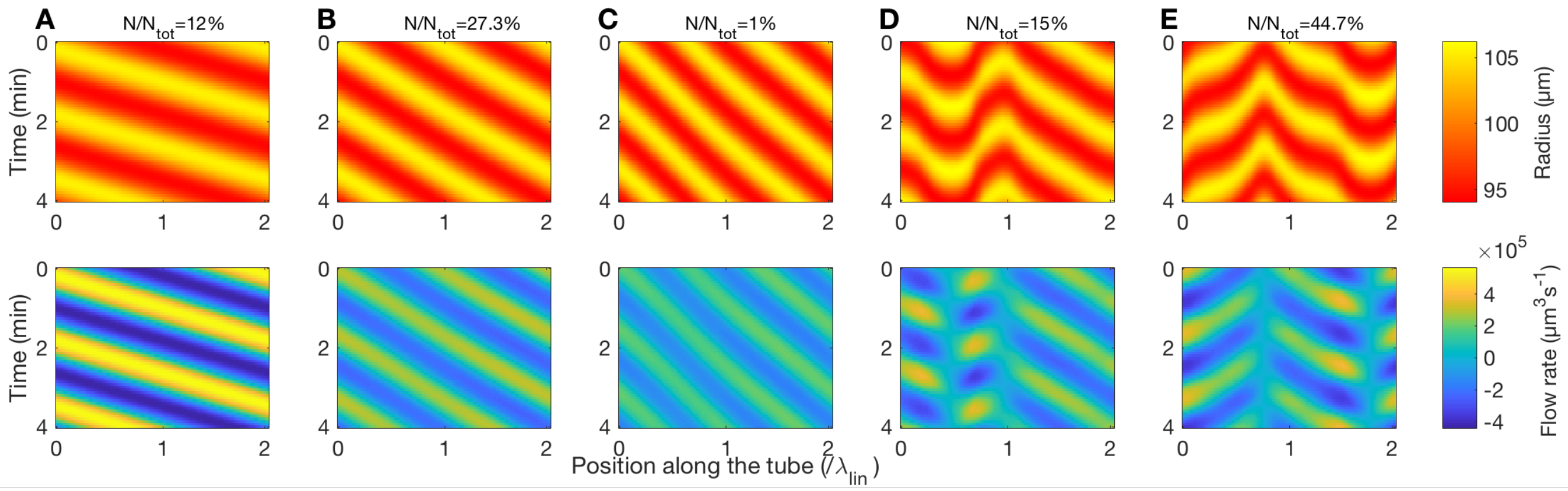}
\caption{Various stable wave patterns in a periodic tube twice as large as the expected intrinsic wavelength $L=2\lambda_{\mathrm{lin}}$. Tube radius (A1-E1) and resulting fluid flow rate (A2-E2) along a tube (horizontal axis) and time (vertical axis). Probability of each pattern in three hundred independent runs denoted above the radius plots.}
\label{fig:patternsExamples}
\end{figure*}
%%%%%%%%%%%%%%%%%%%%%%%%%%%%%%%%%%%%%%%%%%%%%%%%%%%%%%%%%%%%%%%%%%%%%%%%%%%%%%%%
In order to study the self-organization of contractile waves on organisms of varying sizes, we numerically solve model equations \eqref{eq:radius}, \eqref{eq:chemical} in tubes of different lengths $L$ and measure the sizes of the contractile wave patterns $\lambda$, see Fig.~\ref{fig:SimulationNoGrowth}. As model parameters, we choose physiological values for calcium kinetics and acto-myosin cortex mechanics, see Materials and Methods. Tube radius is chosen to match \textit{P. polycephalum} - most renown for scaling contraction waves. To determine the size of the waves, we computed the power spectral density of the radius $a(z,t)$ averaged over ten oscillation periods, and identified the dominant mode. As `wave size' $\lambda$ we denote the inverse of the dominant mode. Note, wave size is not always equivalent to wavelength, in particular if the patterns are antisymmetric (see Fig.~\ref{fig:patternsExamples} B, E, for examples of patterns with different wavelength and equal wave size). Simulations with closed boundary conditions (Fig.~\ref{fig:SimulationNoGrowth} A) fully match our expectations from linear stability analysis, namely waves increasing with tube length up to an upper bound given by the intrinsic wavelength corresponding to the most unstable mode $\lambda_{\mathrm{lin}}$. Surprisingly, simulations with periodic boundary conditions (Fig.~\ref{fig:SimulationNoGrowth} B) also show waves whose sizes do not match $\lambda_{\mathrm{lin}}$, but scale with tube length instead.

Characterizing more precisely the variety of wave patterns, we screen three hundred independent runs with different initial perturbations, for the intermediate tubes length $L=2\lambda_{\mathrm{lin}}$, with periodic boundary conditions, see Fig.~\ref{fig:patternsExamples}, S1. Observed wave patterns can be divided into five cases, by wave size and period-averaged flow rate along the tube (see Fig. S2 for details on the identification of the patterns). There are one, two, or three waves traveling in the same direction (Fig.~\ref{fig:patternsExamples} A, B, and C, respectively), two antisymmetric waves (Fig.~\ref{fig:patternsExamples} D), or two asymmetric waves (Fig.~\ref{fig:patternsExamples} E).
The single wave matching tube length (Fig.~\ref{fig:patternsExamples}A) generates the strongest net fluid flow, exceeding uni-directional multiple wave patterns (Fig.~\ref{fig:patternsExamples} B and C) and asymmetric waves (Fig.~\ref{fig:patternsExamples} E). Patterns with antisymmetric waves (Fig.~\ref{fig:patternsExamples} D) do not create a net flow due to their invariance under space flipping, thus \emph{not} providing any mass transport or long-range mixing. Patterns occur with very different probabilities with almost 50\% resulting in antisymmetric waves with no net flow. The most efficient pattern regarding mixing and transport, where the wave scales with tube length, only has 12\% probability. What measures can make this most efficient pattern more robust? What mechanism drives the scaling of the contractions with the size of the tube?
%%%%%%%%%%%%%%%%%%%%%%%%%%%%%%%%%%%%%%%%%%%%%%%%%%%%%%%%%%%%%%%%%%%%%%%%%%%%%%%%
\subsection*{Growth of the tube leads to the robust scaling of the wave}
%%%%%%%%%%%%%%%%%%%%%%%%%%%%%%%%%%%%%%%%%%%%%%%%%%%%%%%%%%%%%%%%%%%%%%%%%%%%%%%%
\begin{figure}[b!]
\centering
\includegraphics[width=.95\linewidth]{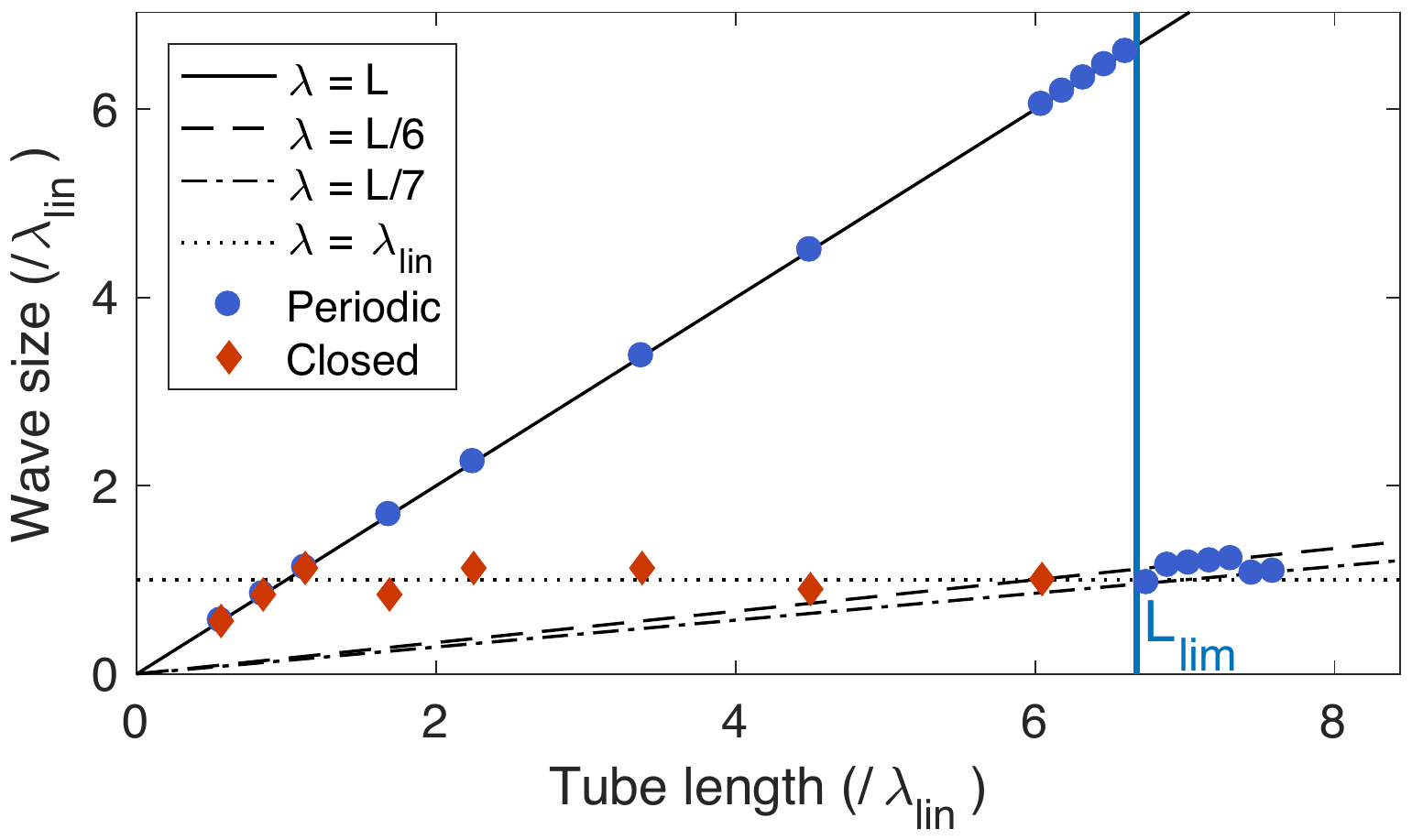}
\centering
\caption{Scaling of the contraction wave in a growing, periodic tube. Wave size $\lambda$ for tubes with periodic (blue circles) and closed (red diamonds) boundaries, grown to different lengths $L$. While wave size in closed tubes saturates at $\lambda_{\mathrm{lin}}$ (dotted line) as predicted from linear stability analysis, waves in tubes with periodic boundaries scale robustly with tube length up to seven-fold the predicted length (blue vertical line). Each data point represents an independent run of a tube grown from initial length of $L=0.2\lambda_{\mathrm{lin}}$.}
\label{fig:SimulationGrowth}
\end{figure}
%%%%%%%%%%%%%%%%%%%%%%%%%%%%%%%%%%%%%%%%%%%%%%%%%%%%%%%%%%%%%%%%%%%%%%%%%%%%%%%%
To investigate robustness and mechanism behind the scaling of contractile waves we performed simulations of growing tubes and measured wave size for periodic or closed boundaries (see Fig.~\ref{fig:SimulationGrowth}). Starting from tubes initially $0.2\lambda_{\mathrm{lin}}$ in length, linear growth is simulated by changing dynamically the mesh size used for spatial discretization. The mesh is refined when the length of the tube doubled. The growth rate is small compared to the contraction period to decouple the dynamics of the system from growth. After the tube reached its target length, simulations were continued for roughly $200$ additional contraction periods to ensure that growth has no impact on the simulated pattern.

%%%%%%%%%%%%%%%%%%%%%%%%%%%%%%%%%%%%%%%%%%%%%%%%%%%%%%%%%%%%%%%%%%%%%%%%%%%%%%%%
\begin{figure}[t]
\centering
\includegraphics[width=.95\linewidth]{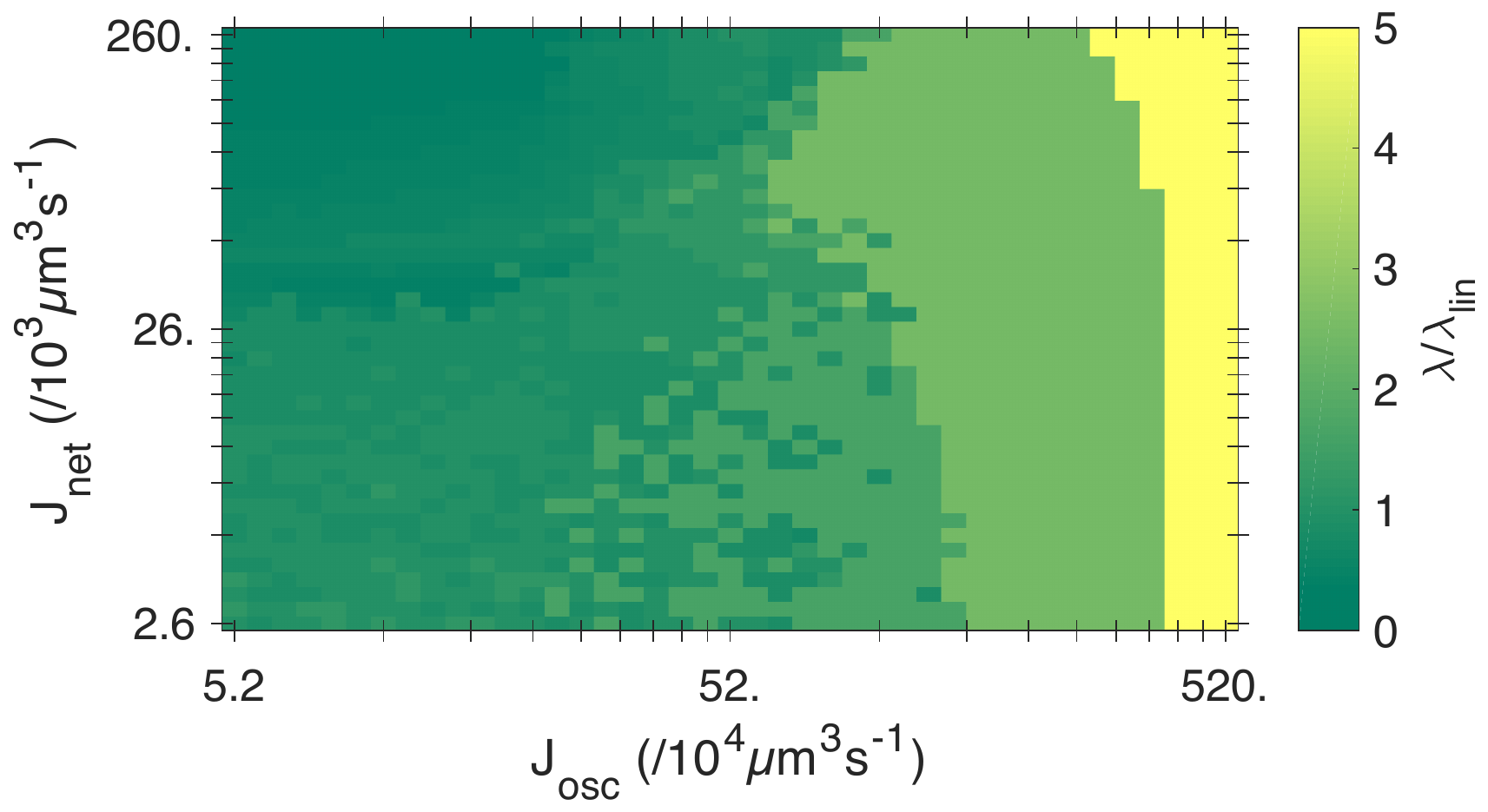}
\centering
\caption{Net flow and amplitude of oscillatory flow have competing effects
on the wave size. Phase map of wave sizes in tubes with an imposed flow $J=J_{\mathrm{net}}+J_{\mathrm{osc}}\cos(\omega t)$ at one boundary. Full scaling arises at high oscillatory flow $J_{\mathrm{osc}}$ almost independent of net flow $J_{\mathrm{net}}$.}
\label{fig:imposedFlow}
\end{figure}
%%%%%%%%%%%%%%%%%%%%%%%%%%%%%%%%%%%%%%%%%%%%%%%%%%%%%%%%%%%%%%%%%%%%%%%%%%%%%%%%
In agreement with linear stability analysis we find waves in tubes with closed boundaries grow with tube length only up to the upper bound $\lambda_{\mathrm{lin}}$. Yet, for periodic boundary conditions, waves scales with the length of the tube up to the seven-fold $\lambda_{\mathrm{lin}}$, see Fig.~\ref{fig:SimulationGrowth}. Above this limit to the scaling $L_{\mathrm{lim}}$, the wave multiplies into six or seven smaller waves, whose size matches roughly $\lambda_{\mathrm{lin}}$ (Fig.~\ref{fig:SimulationGrowth}). Results are robust against variations in parameters. Particularly, changing the fluid viscosity $\mu$ varies the scaling limit $L_{\mathrm{lim}}$ and the factor of mode-multiplication  $n=L_{\mathrm{lim}}/\lambda_{\mathrm{lin}}$. Contrary to previous reaction-diffusion systems capable of mode-doubling or tripling when simulated on growing domains \cite{crampin_mode-doubling_2002}, many values of $n$ are accessible, see Fig. S3 for $n\in[5,8]$. From Eq.~\ref{eq:most unstable wave number}, we can see that the predicted wavelength scales like $\lambda_{\mathrm{lin}}\propto\mu^{-1/4}$. On the other hand, a dimensional analysis of the advective term in Eq.~\ref{eq:chemical} leads to a typical scale proportional to $\mu^{-1/2}$, consistent with our simulations showing $L_{\mathrm{lim}}\propto\mu^{-0.51}$ (see Fig. S3). As $\lambda_{\mathrm{lin}}$ and $L_{\mathrm{lim}}$ scale differently with viscosity, the mode-multiplication factor $n$ changes accordingly.
Noteworthy, decreasing the viscosity increased the scaling limit. A lower viscosity does not change any mechanical properties of the tube but increases the flow velocity, and thus advection of the contraction-triggering chemical. This suggests that flow-driven transport is crucial for the observed scaling mechanism and the upper scaling limit.
%%%%%%%%%%%%%%%%%%%%%%%%%%%%%%%%%%%%%%%%%%%%%%%%%%%%%%%%%%%%%%%%%%%%%%%%%%%%%%%%
\subsection*{Scaling of the wave is due to oscillatory flows}
%%%%%%%%%%%%%%%%%%%%%%%%%%%%%%%%%%%%%%%%%%%%%%%%%%%%%%%%%%%%%%%%%%%%%%%%%%%%%%%%
In order to distinguish the role of net flow $J_\mathrm{net}$ and oscillatory flow $J_\mathrm{osc}$ in establishing the scaling, we investigated dynamics in tubes with an imposed inflow on one end of the tube. The flow is set as $J=J_{\mathrm{net}}+J_{\mathrm{osc}}\cos(\omega t)$. To limit our study to a two dimensional parameter space, we set $\omega$ to the natural angular frequency of our system, calculated using linear stability analysis, Eq. S1. The values of $J_\mathrm{net}$ and $J_\mathrm{osc}$ were chosen to be comparable with the values generated spontaneously in simulations of periodic tubes (see Fig. S4 for a comparison of $J_\mathrm{net}$ and $J_\mathrm{osc}$ with the flows measured in the simulations of Fig.~\ref{fig:SimulationGrowth}).

Imposed flow boundary conditions result in long-ranged contraction patterns, see Fig.~\ref{fig:imposedFlow}. Interestingly, net flow $J_\mathrm{net}$ and $J_\mathrm{osc}$ have competing effects on the observed wave size. Contrary to our expectations, the net flow $J_{\mathrm{net}}$ has little impact on contraction wave size. Although small, its effect depends on the oscillatory flow. $J_{\mathrm{net}}$ decreases or increases wave size for low and large values of $J_{\mathrm{osc}}$, respectively. The oscillatory part of the flow $J_{\mathrm{osc}}$, on the other hand, increases sharply the wavelength for any value of $J_{\mathrm{net}}$. Interestingly, the time necessary to establish a stable pattern of contractions is shorter as the wave size grows (see Fig. S4). Thus, we find that oscillating flow, rather than net flow, is the key to scaling with system size much beyond the intrinsic length scale of the instability.
%%%%%%%%%%%%%%%%%%%%%%%%%%%%%%%%%%%%%%%%%%%%%%%%%%%%%%%%%%%%%%%%%%%%%%%%%%%%%%%%
\section*{Discussion}
%%%%%%%%%%%%%%%%%%%%%%%%%%%%%%%%%%%%%%%%%%%%%%%%%%%%%%%%%%%%%%%%%%%%%%%%%%%%%%%%
We have studied the self-organization of long-ranged fluid flows in tubular-shaped cells, due to the coupling of cortex contractions to an advected, contraction-triggering chemical. Our minimal two-component model system describing cortex and chemical dynamics predicts self-sustained contraction waves of wave size $\lambda_{\mathrm{lin}}$. Numerical simulations of the model confirm these predictions in tubes with closed boundaries. Yet, in tubes with periodic boundary conditions we find flows to scale with tube length much beyond the predicted wave size $\lambda_{\mathrm{lin}}$. Robust scaling is observed when tubes are slowly grown longer than $\lambda_{\mathrm{lin}}$, ``mode-multiplying'' at a scaling limit $L_{\mathrm{lim}}=n\lambda_{\mathrm{lin}}$, $n=6.6$.
Simulations of fluids with different viscosities and tubes with imposed inflow show that the oscillatory flow is the key behind this unexpected scaling on such long length scales.

From a dynamical systems point of view, the observation of mode-multiplying at factors of up to eight vastly exceeds previous observations of mode-doubling or mode-tripling \cite{crampin_mode-doubling_2002}. Within our model we find that mode-multiplying is determined by the ratio between the scaling limit $L_{\mathrm{lim}}$ and the linearly unstable wavelength $\lambda_{\mathrm{lin}}$. The impact of the viscosity of the cytoplasm, as an example, was investigated, and the different scaling with viscosity, $\lambda_{\mathrm{lin}}\propto\mu^{-1/4}$ and $L_{\mathrm{lim}}\propto\mu^{-1/2}$, explained the splitting of the contraction wave to different modes.

The oscillatory nature of fluid flows allowed by periodic boundary conditions or by imposed flow is crucial to generate scaling beyond the intrinsic wavelength $\lambda_{\mathrm{lin}}$, Eq.~\ref{eq:most unstable wave number}. The value of the intrinsic wavelength may vary broadly between different systems. Assuming our representative parameter values and only taking into account that cell elastic modulus and cortex viscosity scale with cortex thickness $h$ over cell radius $a^{*}$, a rule of thumb for the intrinsic wavelength is $\lambda_{\mathrm{lin}}\simeq\SI{1}{m^{\frac{1}{2}}}(h a^{*})^{\frac{1}{4}}$. This rule of thumb implies that only based on the intrinsic wavelength, a doubling in system size would requires a 16-fold increase in radius to allow intrinsic wavelength to match doubled system size. Alternatively, oscillatory flow boundary conditions are required to allow for scaling with system size beyond the intrinsic wavelength as the system grows.
Given the possible range of biological parameters, this rule of thumb sets a scale for the radius over which long-ranged scaling due to oscillatory flows is likely to be at work. Note, that the scale predicted here can only be a rough estimate as our mechanochemical model only accounts for the role of calcium. In fact, additional cortex regulation machinery might be important in a specific system, as for example in zebrafish primordial germ cells, where the SDF-1 protein \cite{blaser_migration_2006} leads to the accumulation of free calcium then triggering acto-myosin contraction. Moreover, measurements of the mechanical and geometrical properties of the cell cortex show large variations and increase further the uncertainty.
The key insight is that oscillatory fluid flows can generate scaling contraction waves and that it may be worth checking for their role in large systems exceeding the intrinsic wavelength.

For the system best studied for its cortex driven cytoplasmic flows, \textit{Physarum polycephalum}, the predicted wave size is $\lambda_{\mathrm{lin}}=\SI{7.1}{mm}$, about an order of magnitude smaller than the coordinated contraction waves observed on scales of up $\SI{2}{\centi\meter}$ \cite{alim_random_2013}, but well within the range of $L_{\mathrm{lim}}=\SI{4.7}{\centi\meter}$, in agreement with our predictions. At these large scales \textit{P. polycephalum} forms a network of tubes with more viscous bags pooling fluid at the growing fronts. It is fascinating to speculate how the network morphology impacts the dynamics of contractile waves. It is likely that the contractions of the viscous bags at the growing fronts here do serve as pumps very much similar to the imposed flow boundary conditions we implemented. The growing fronts could thereby also account for the resurrection of scaling contraction waves after contraction stopped due to harmful external stimuli \cite{Bauerle:2017}. In contrast to \textit{P. polycephalum} to date detailed quantitative data is lacking in other systems to allow for quantitative comparison. Yet, cortex contractions and oscillatory flows are very general component for many other systems, even beyond the single cell. Thus, the interplay of fluid flows and mechanical oscillations resulting in scaling might be broadly relevant.

In general, our model broadens the budding understanding of the fundamental role of cytoplasmic flows in a large class of biophysical systems \cite{woodhouse_cytoplasmic_2013,goldstein_physical_2015,monteith_mechanism_2016,quinlan_cytoplasmic_2016}. In very diverse systems, flows appear to be a fundamental part of a self-organized machinery. 
In our case, their oscillations are crucial to drive and organize patterns of contractions on a large scale, a mechanism likely present in many other biological systems.
More fundamentally, our result opens perspectives of how including active advection in classical reaction-diffusion frameworks leads to unexpected observations such as scaling.
%%%%%%%%%%%%%%%%%%%%%%%%%%%%%%%%%%%%%%%%%%%%%%%%%%%%%%%%%%%%%%%%%%%%%%%%%%%%%%%%
\section{Methods}
\subsubsection*{Implementation}
Numerical solutions of the model equation were explored with a custom written Crank-Nicholson scheme implemented in MATLAB (The Mathworks). Simulations started from the spatially uniform equilibrium value for tube radius and chemical concentration. To perturb the stable state, uncorrelated, Gaussian fluctuations of standard deviation 0.01 were added to the radius.  Patterns of the contractions appear in a few periods, and stabilize completely within 50 periods. Three kinds of boundary conditions were implemented: periodic, closed, or flow. For closed boundary conditions, the radius and the chemical concentration at the boundaries of the tube are both assumed to be equal to their value at the uniform equilibrium. For flow boundary conditions, fluid flow is imposed on one end of the tube in flow boundary conditions.

\subsubsection*{Parameters}
Simulations parameters were $\mu=\SI{1.5d-3}{\pascal.\second}$ for the viscosity of the cytoplasm \cite{puchkov_intracellular_2013}, $a^{*}=\SI{100}{\micro\meter}$ for the radius of the viscoelastic tube, $E=\SI{10}{\pascal}$ for its effective stiffness (assuming a Young's modulus of $\SI{100}{\pascal}$ \cite{naib-majani_morphology_1988,salbreux_actin_2012} and a thickness of the tubes of $h=a^{*}/10$), $\eta=E\times24s$ for its effective viscosity \cite{naib-majani_morphology_1988,feneberg_dictyostelium_2001,saha_determining_2016}, $\kappa=100E$ for the non-linear elasticity, $\sigma_0=3E$ for the active stress \cite{salbreux_actin_2012,maitre_pulsatile_2015}, $D=\SI{3.33d-10}{\meter^{2}.\second^{-1}}$ \cite{donahue_free_1987,radszuweit_active_2014} for the diffusion of the tension activator in the cytoplasm, $a^*c^*/(2p_c)= a^*/(2d_c)=\SI{96}{s}$ \cite{glogauer_calcium_1997,lee_regulation_1999} for the time-scale of its regulation, $\epsilon_{c}=0.3$ for the threshold of its stretch-activated supply, and $\epsilon_{\sigma}=2$ for the stretch-inhibition of the active stress. Other parameters were eliminated in the non-dimensional equations. The mechanical parameters, particularly $\kappa$ and $\epsilon_\sigma$, were tuned to lead to deformations of about ten percent, typical for \textit{P. polycephalum} \cite{alim_random_2013}. The resulting flow velocities in our simulations were typically $\bar{u}=$\SIrange{10}{30}{\micro\meter.\second^{-1}}, matching cytoplasmic flows for \textit{P. polycephalum} \cite{alim_mechanism_2017}.

\subsection*{Acknowledgment}
This work was supported by the Max Planck Society.

% Bibliography
\bibliographystyle{pnas-new}
\bibliography{bibJD,bibKA}

\end{document}